\documentclass[aps,pre,twocolumn,showpacs]{revtex4-1}

\usepackage{amsmath, amssymb}
\usepackage{epsfig}
\usepackage{braket}

\newcommand{\prt}[2]{\frac{\partial #1}{\partial #2}}
\newcommand{\prts}[3]{\frac{\partial^{#3} #1}{\partial {#2}^{#3}}}

\newcommand{\tod}{\stackrel{\mathrm{d}}{\to}}

\newcommand{\mr}[1]{\mathrm{#1}}
\renewcommand{\eqref}[1]{Eq.~(\ref{#1})}
\newcommand{\eqsref}[1]{Eqs.~(\ref{#1})}
\newcommand{\pref}[1]{(\ref{#1})}
\newcommand{\figref}[1]{Fig.~\ref{#1}}

\newcommand{\secref}[1]{Sec.~\ref{#1}}

\begin{document}

\title{When fast and slow interfaces grow together:\\ connection to the half-space problem of the Kardar-Parisi-Zhang class}
\author{Yasufumi Ito}
\author{Kazumasa A. Takeuchi}
\email{kat@kaztake.org}
\affiliation{Department of Physics, Tokyo Institute of Technology,
 2-12-1 Ookayama, Meguro-ku, Tokyo, 152-8551, Japan.}

\date{\today}

\begin{abstract}
 We study height fluctuations of interfaces in the $(1+1)$-dimensional Kardar-Parisi-Zhang (KPZ) class, growing at different speeds in the left half and the right half of space. Carrying out simulations of the discrete polynuclear growth model with two different growth rates, combined with the standard setting for the droplet, flat, and stationary geometries, we find that the fluctuation properties at and near the boundary are described by the KPZ half-space problem developed in the theoretical literature. In particular, in the droplet case, the distribution at the boundary is given by the largest-eigenvalue distribution of random matrices in the Gaussian symplectic ensemble, often called the GSE Tracy-Widom distribution. We also characterize crossover from the full-space statistics to the half-space one, which arises when the difference between the two growth speeds is small.
\end{abstract}


\maketitle


\section{Introduction}

Interface growth and resulting scale-invariant fluctuations
have been an important target of non-equilibrium physics
for decades \cite{barabasi},
but they began to take a unique position
when the paradigmatic universality class in this context,
namely the Kardar-Parisi-Zhang (KPZ) class,
turned out to be tractable by exact solutions in one dimension
\cite{Kriecherbauer2010,Corwin2012,HalpinHealy.Takeuchi-JSP2015,Takeuchi-a2017}.
Suppose an interface grows upward
on a one-dimensional substrate,
then the growth can be described in terms of its height profile $h(x,t)$
at spanwise position $x$ and time $t$.
If this interface belongs to the KPZ class,
$h(x,t)$ is known to grow as
\begin{equation}
 h(x,t)\simeq v_{\infty}t+(\Gamma t)^{1/3}\chi(X,t) \label{eq:height-KPZ}
\end{equation}
with a rescaled coordinate $X:=cx/t^{2/3}$,
non-universal coefficients $v_{\infty},\Gamma,c$,
and a rescaled random variable $\chi(X,t)$
that represents the height fluctuations.
The exponent values $1/3$ and $2/3$ in those equations
characterize the $(1+1)$-dimensional KPZ class
\cite{PhysRevLett.56.889,Forster.etal-PRA1977}.
The modern developments triggered by exact studies are more concerned
with finer fluctuation properties of $\chi(X,t)$,
such as its distribution function and correlation properties
\cite{Kriecherbauer2010,Corwin2012,HalpinHealy.Takeuchi-JSP2015,Takeuchi-a2017}.
They are also universal
and were indeed identified in experiments of liquid-crystal turbulence
\cite{PhysRevLett.104.230601,*Takeuchi.etal-SR2011,*Takeuchi2012,*PhysRevLett.119.030602}.

Among important outcomes of the recent developments
\cite{Kriecherbauer2010,Corwin2012,HalpinHealy.Takeuchi-JSP2015,Takeuchi-a2017},
particularly noteworthy are the facts that
(1) the universal fluctuation properties of $\chi(X,t)$ can be classified
according to the interface geometry \cite{Prahofer2000},
or equivalently the initial condition, and (2) in prototypical cases,
a connection to random matrix theory \cite{Mehta-Book2004,*rmtbook} was found
\cite{Sasamoto-JSM2007}.
Specifically, if an interface grows from a single nucleus
--~hereafter referred to as the droplet geometry~--,
the asymptotic distribution is given by that of the largest eigenvalue
of random matrices in the Gaussian unitary ensemble (GUE),
called the GUE Tracy-Widom (GUE-TW) distribution \cite{Tracy1994,*Tracy1996}.
For interfaces growing from a flat substrate,
the TW distribution for the Gaussian orthogonal ensemble (GOE) arises.
The asymptotic distribution was also obtained for the stationary case,
i.e., with the initial condition drawn from the stationary measure,
which is then given by the Baik-Rains (BR) distribution
\cite{Baik2000,Prahofer2000}.
These three constitute the representative cases, sometimes called
universality subclasses of the $(1+1)$-dimensional KPZ class.
Two of them are related to prominent ensembles of random matrix theory
\cite{Mehta-Book2004,*rmtbook}.

One may then wonder if the TW distribution of the other,
equally established ensemble of random matrices,
namely the Gaussian symplectic ensemble (GSE) \cite{Mehta-Book2004},
can arise in the KPZ class.
The answer is yes; it was theoretically found
for several semi-infinite systems with the droplet geometry
\cite{Sasamoto-JSM2007,Baik2001symmetrized,Prahofer2000,Sasamoto2004,Gueudre2012,Baik.etal-a2016,*Baik2017,Barraquand.etal-a2018},
where $h(x,t)$ is defined with $x\geq0$
and the boundary at $x=0$ is either constrained by some condition
or driven with a different model parameter.
To give examples,
it was shown \cite{Baik2001symmetrized,Prahofer2000,Sasamoto2004} that,
the polynuclear growth (PNG) model with a different nucleation rate
at the origin exhibits the GSE-TW and Gaussian distributions
for small and large growth rates, respectively,
and the GOE-TW distribution at the critical point.
The GSE-TW distribution was also derived for the KPZ equation
with an absorbing wall at the origin \cite{Gueudre2012},
and it was argued that the same conclusion should hold if $\partial_x{}h\geq0$
at the origin.
Such a half-space problem has also been studied
for the flat and stationary geometries
\cite{Baik2001symmetrized,PrahoferBook2002}.
However, from the experimental viewpoint,
controlling the growth rate or the interface slope at the boundary
is unrealistic in many cases.
As a result, the GSE-TW distribution, as well as other universal properties
predicted for the half-space problems, still remain experimentally elusive.

In this work, we propose a more realistic situation
to study the half-space problem, where an interface grows
in both $x<0$ and $x\geq0$, but at different speeds in the two regions.
We implement this ``biregional'' setting numerically,
using the discrete PNG model
with the droplet, flat, and stationary geometries, and find
the characteristic properties of the corresponding half-space problems.
In particular, the GSE-TW distribution was found in the droplet case,
as well as the associated spatial correlation near the boundary.
If the difference between the two growth speeds is small,
crossover from the usual full-space statistics to the half-space one is found.
We show how this crossover is controlled by the growth speed difference.


\section{Model}

\begin{figure}[t]
 \centering
 \includegraphics[width=0.9\columnwidth]{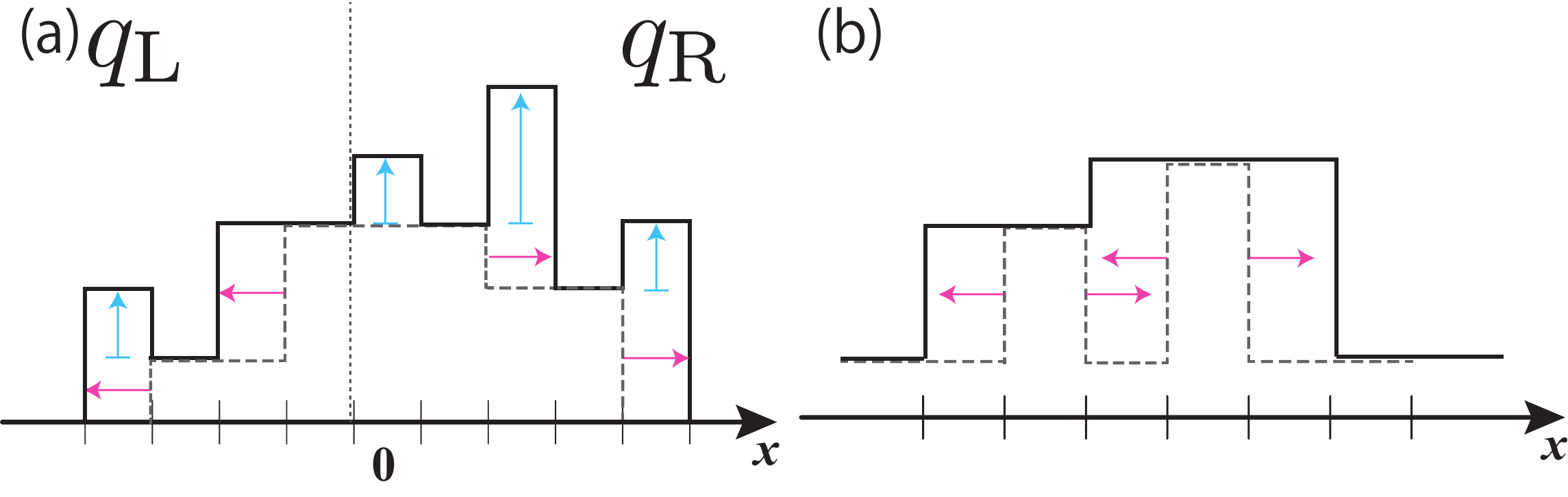}
 \caption{(Color online).
  Sketch of the time evolution rules of the discrete PNG model.
  (a) An example of the interface evolution
  from time $t$ (dashed line) to $t+1$ (solid line).
  The vertical arrows indicate elevation by random nucleations
  and the horizontal arrows show plateau expansion.
  The growth parameter is $q=q_\mr{L}$ for $x<0$ and $q=q_\mr{R}$ for $x\geq0$.
  (b) When two plateaus encounter, the higher one overrides.}
 \label{fig:PNGdefi}
\end{figure}

We use the discrete PNG model
and adapt it for our biregional setting.
In the following, $x\in\mathbb{Z},\,t\in\mathbb{N}_0,\,h(x,t)\in\mathbb{N}_0$.
The initial condition is $h(x,0)=0$.
Time evolution is illustrated in Fig.~\ref{fig:PNGdefi}.
Briefly, random nucleation occurs locally,
which increases $h(x,t)$ at the nucleation point
by a random integer $\omega(x,t)$, and the produced projection expands
laterally at unit speed [Fig.~\ref{fig:PNGdefi}(a)].
When low and high plateaus encounter, the higher one overrides
 [Fig.~\ref{fig:PNGdefi}(b)].
Those evolution rules are expressed by
\begin{equation}
 h(x,t+1)=\max\{h(x-1,t),h(x,t),h(x+1,t)\} + \omega(x,t+1).  \label{eq:PNGrule}
\end{equation}
Here, following Ref.~\cite{Sasamoto2004}, we consider the case in which
nucleation can occur only at even (resp. odd) sites at even (resp. odd) times.
If nucleation is allowed, $\omega(x,t)$ is drawn independently
from the geometric distribution with parameter $0\leq{}q<1$,
set to be $q=q_\mr{L}$ for $x<0$ and $q=q_\mr{R}$ for $x\geq0$.
More explicitly, with $k\in\mathbb{N}_0$,
\begin{equation}
 \mathrm{Prob}[\omega(x,t)=k]=
 \begin{cases}
  (1-q_{\mathrm{L}})q_{\mathrm{L}}^k,   & (x<0),   \\
  (1-q_{\mathrm{R}})q_{\mathrm{R}}^{k}, & (x\ge0), 
 \end{cases} \label{eq:omega}
\end{equation}
if $x-t$ is odd.
Otherwise $\omega(x,t)=0$.

The advantage of using such an alternating update is that
the scaling coefficients $v_{\infty},\Gamma,c$ in \eqref{eq:height-KPZ}
are known analytically as follows, in the case of the homogeneous growth
$q=q_\mr{L}=q_\mr{R}$ \cite{Sasamoto2004}:
\begin{equation}
 v_{\infty} = \frac{\sqrt{q}}{1-\!\sqrt{q}}, \quad
 \Gamma = \frac{\sqrt{q}(1+\!\sqrt{q})}{2(1-\!\sqrt{q})^3}, \quad
 c = \frac{q^{1/6}}{2^{1/3}(1+\!\sqrt{q})^{2/3}}.  \label{eq:PNGparam}
\end{equation}
Using these coefficients, we can define the rescaled height by
\begin{equation}
 H(X,t)
 := \frac{h(x=Xt^{2/3}/c,t)-v_{\infty}t}{(\Gamma t)^{1/3}}
 \simeq \chi(X,t).  \label{eq:HeightRescaling}
\end{equation}
As we explain below, even if $q_\mr{L}\neq{}q_\mr{R}$,
the same expressions remain valid in the region with the larger $q$.
In the following, we set $q_\mr{L}\leq{}q_\mr{R}$
(growth is faster in $x\geq0$) without loss of generality.

Now we describe how we implement the droplet, flat, and stationary geometries
in this model.

\subsection{Droplet}  \label{sec:Droplet}

Following the standard method for the PNG model \cite{Prahofer2000},
we realize the droplet geometry by restricting nucleations to $|x|\leq{}t$
(in addition to the alternating rule).
Thereby the growth process starts at the origin,
forming a circular interface in the homogeneous case $q_\mr{L}=q_\mr{R}$.
If $q_\mr{L}<q_\mr{R}$, we obtain a deformed interface.
This is what we call the droplet case.

\subsection{Flat}

In the flat case, nucleations can occur at any sites with $x-t$ even.
Therefore, for simulations, the system boundary must be explicitly considered.
Here we use
$h(\pm(L+1),t)=0$ in \eqref{eq:PNGrule}.
Since we are interested in statistical properties at $x=0$ and nearby,
the choice of the boundary condition has little influence
as long as $L$ is sufficiently large.
Here we use $L=t_\mr{max}+2$,
where $t_\mr{max}=10^4$ is the final time of the simulations.

\subsection{Stationary}

Here the stationary geometry refers to the case where
the initial condition consists of a pair of stationary interfaces
in the two regions, connected at the boundary.
If $q_\mr{L}<q_\mr{R}$, the mismatch of the growth speeds
makes the interface non-stationary.
Nevertheless, we use the term stationary, because
the interface shows characteristics of the stationary interfaces,
e.g., the BR distribution, far from the boundary.

While the initial condition described above might be directly implemented,
to avoid the boundary effect,
here we adopt the method used in Ref.~\cite{Prahofer2000}.
Specifically, we take the droplet geometry described in \secref{sec:Droplet}
and add an additional nucleation term $\omega_\pm(x,t+1)$ to \eqref{eq:PNGrule}
at the droplet edges $x=\pm t$.
The edge nucleation also follows the geometric distribution \pref{eq:omega}
with parameter $q_\pm$, which is set to be
$q_{+}=q_{\mathrm{R}}^{1/2}$ and $q_{-}=q_{\mathrm{L}}^{1/2}$.
Those values are chosen so that the generated interface
is indeed in the stationary state as defined above
\cite{Prahofer2000,Takeuchi-a2017}.

\subsection{Limiting cases}  \label{sec:limit}

Clearly, if $q_\mr{L}=q_\mr{R}$,
our system becomes the standard discrete PNG model.
The asymptotic fluctuation properties are therefore exactly known
\cite{Prahofer2000}.
For the one-point distribution, it is
\begin{equation}
 H(X,t)_{q_\mr{L}=q_\mr{R}} \tod \begin{cases}
  \chi_\mr{GUE} - X^2                     & (\text{droplet}),    \\
  2^{-2/3}\chi_\mr{GOE} =: \chi'_\mr{GOE} & (\text{flat}),       \\
  \chi_\mr{BR}                            & (\text{stationary}), 
 \end{cases}  \label{eq:limit1}
\end{equation}
where $\chi_\mr{GUE},\,\chi_\mr{GOE},\,\chi_\mr{BR}$
are the standard random variables
of the GUE-TW, GOE-TW, BR distributions
\cite{Tracy1994,Tracy1996,Baik2000}, respectively,
and ``$\tod$'' denotes convergence in distribution.

In the other limiting case $q_\mr{L}=0$,
our system becomes equivalent to the half-space PNG model
\cite{Sasamoto2004}
without boundary nucleation.
With other theoretically solid results for the half-space problem
\cite{Baik2001symmetrized,PrahoferBook2002},
the one-point distribution at the origin is
\begin{equation}
 H(0,t)_{q_\mr{L}=0} \tod \begin{cases}
  2^{1/2}\chi_\mr{GSE} =: \chi_\mr{GSE}' & (\text{droplet}),    \\
  \chi_\mr{GUE}                          & (\text{flat}),       \\
  \chi_\mr{GOE}                          & (\text{stationary}), 
 \end{cases}  \label{eq:limit2}
\end{equation}
with the random variable $\chi_\mr{GSE}$ of the GSE-TW distribution
\cite{Tracy1996}.

Between those limiting cases, $0<q_\mr{L}<q_\mr{R}$,
we have the situation where the interface grows at different speeds.
This is the primary target of the present paper.


\section{results}

In the following we fix $q_\mr{R} = 0.25$,
and $q_\mr{L}$ is varied in the range $0\leq{}q_\mr{L}\leq{}q_\mr{R}$.
We carried out simulations for the three geometries described above.
Statistical results were obtained from 100,000 realizations
for each case, unless otherwise stipulated.


\subsection{Height Profile}

\begin{figure}[t]
 \centering
 \includegraphics[width=\columnwidth]{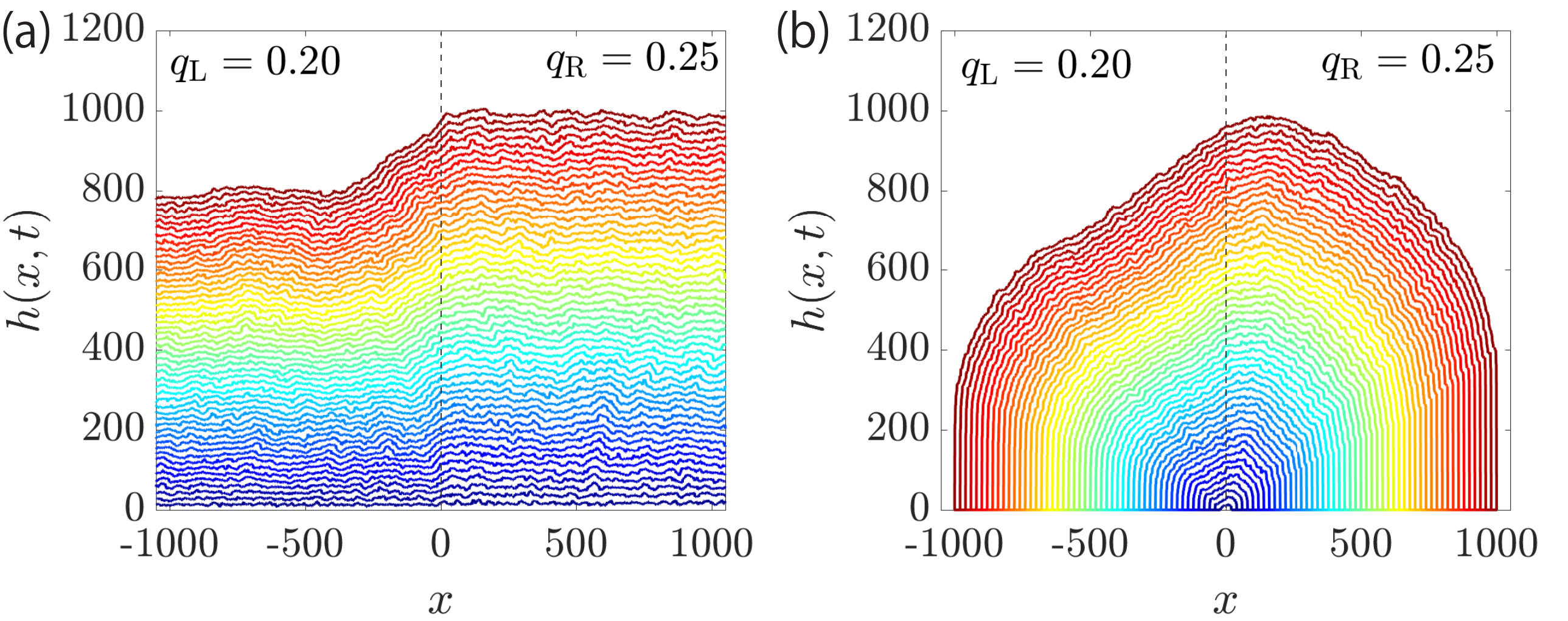}
 \caption{(Color online).
  An example of interfaces in the flat (a) and circular (b) geometries,
  with $q_{\mathrm{L}}=0.2$ and $q_{\mathrm{R}}=0.25$.
  The height profiles recorded every 20 time steps are shown.
 }
 \label{fig:interface}
\end{figure}

Typical height profiles for the flat and droplet cases
are shown in Fig.~\ref{fig:interface}.
In the flat case,
the interface consists of two flat regions growing at different speeds,
connected by a slope near the boundary [Fig.~\ref{fig:interface}(a)].
Interestingly, the slope is found to be kept constant in time, though
the difference between the heights far from the boundary increases.
This can be understood by considering
the noiseless version of the KPZ equation:
\begin{equation}
 \begin{cases}
  \prt{h}{t} = \nu\prts{h}{x}{2} + \frac{\lambda}{2}\left(\prt{h}{x}\right)^2 + v_\mr{L},
   & (x<0),      \\
  \prt{h}{t} = \nu\prts{h}{x}{2} + \frac{\lambda}{2}\left(\prt{h}{x}\right)^2 + v_\mr{R},
   & (x \geq 0), 
 \end{cases}  \label{eq:NoiselessKPZ1}
\end{equation}
where $\nu$ and $\lambda$ are constant coefficients
and $v_\mr{L}<v_\mr{R}$ denote the two growth speeds.
The asymptotic solution $h_\mr{asymp}(x,t)$ is
\begin{equation}
 h_\mr{asymp}(x,t) = \begin{cases}
  v_\mr{R}t + \sqrt{\frac{2\Delta v}{\lambda}}x,                                                              & (x<0),      \\
  v_\mr{R}t +\frac{2\nu}{\lambda} \log\left( 1+\frac{\lambda}{2\nu}\sqrt{\frac{2\Delta v}{\lambda}}x \right), & (x \geq 0), 
 \end{cases}  \label{eq:NoiselessKPZ2}
\end{equation}
with $\Delta{}v:={}v_\mr{R}-v_\mr{L}$.
This accounts for the numerically observed appearance of the constant slope,
which penetrates into the faster-growth region over a finite distance.

In the droplet case,
the asymptotic mean profile
in the homogeneous growth condition $q_{\mr{L}}=q_{\mr{R}}=q$
is known to be \cite{Sasamoto2004,Rost1981}
\begin{equation}
 h(x,t)\simeq v_{\infty}t\frac{\sqrt{q}+\sqrt{1-(x/t)^2}}{1+\sqrt{q}},
 \label{eq:limit-shape}
\end{equation}
i.e., an expanding semicircle with a rising center.
Then, in our biregional setting $q_\mr{L}<q_\mr{R}$,
if there were no interaction between the two regions,
two quadrants of different radii would grow.
However, the same sort of interaction as for the flat case exists,
producing a similar intermediate region
of a constant slope [Fig.~\ref{fig:interface}(b)].



\subsection{Distribution at the Boundary}  \label{sec:OnePoint}

Now we study how the interface fluctuates around the mean profile,
at and near the boundary.
The result of the mean height profile, in particular \eqref{eq:NoiselessKPZ2},
suggests that this boundary region is essentially controlled
by the faster-growth region.
Therefore, the height should be rescaled by \eqref{eq:HeightRescaling}
with \eqref{eq:PNGparam} and $q=q_\mr{R}=0.25$;
specifically, $v_\infty=1,\,\Gamma=3,\,c=3^{-2/3}$.
In this section, we study the rescaled height fluctuations at the boundary,
$H(0,t)$.

\begin{figure}[t]
 \centering
 \includegraphics[width=\columnwidth]{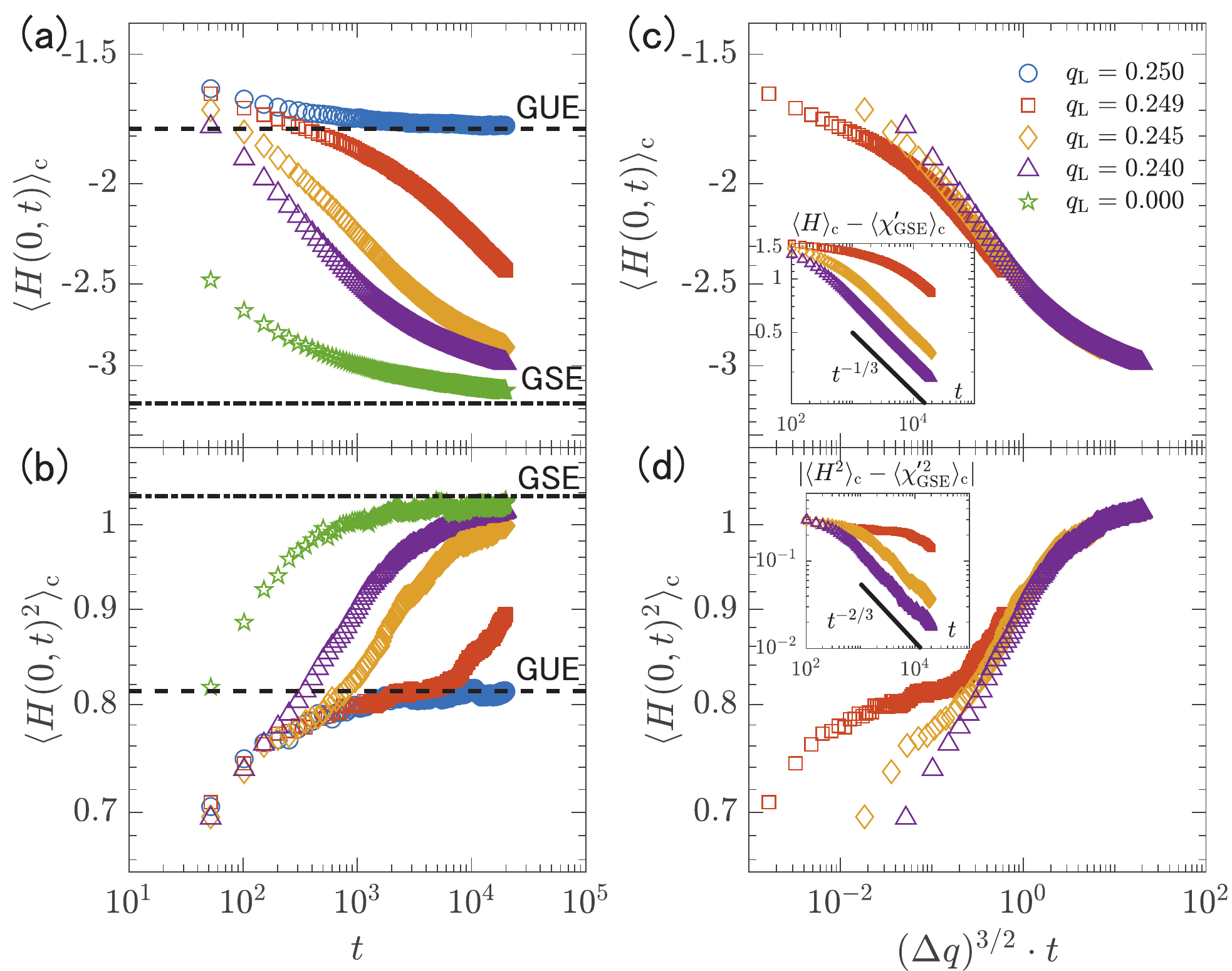}
 \caption{
  The mean and the variance of the rescaled height $H(0,t)$ at the boundary
  for the droplet case.
  Different colors and symbols correspond to different values of $q_{\mr{L}}$,
  as shown in the legend of panel (c).
  The horizontal lines indicate the mean and the variance of
  $\chi_\mr{GUE}$ (dashed) and $\chi_\mr{GSE}'$ (dash-dot).
  The raw data in (a)(b) are plotted against $(\Delta{}q)^{3/2}t$
  in (c)(d).
  The insets show approach to the GSE-TW values.
  To improve statistical accuracy, the data for $q_\mr{L}=0.240$ were
  obtained from 500,000 realizations.
  In view of the alternating character of the updating,
  only data at even times are shown.
 }
 \label{fig:droplet-cumulant}
\end{figure}

First, the results for the droplet case are shown
in Fig.~\ref{fig:droplet-cumulant}.
Figures~\ref{fig:droplet-cumulant}(a) and (b) show the mean $\braket{H(0,t)}$
and the variance $\braket{H(0,t)^2}_\mr{c}$, respectively,
with varying $q_{\mathrm{L}}$.
For the two limiting cases discussed in \secref{sec:limit},
i.e., for the homogeneous case $q_{\mathrm{L}}=q_{\mathrm{R}}=0.250$
(blue circles) and the half-space case $q_\mr{L}=0$ (green stars),
our numerical data support the expected convergence
to the GUE-TW and GSE-TW distributions, respectively
 [\eqsref{eq:limit1} and \pref{eq:limit2}].
The data in between correspond to the results of our biregional setting,
which seem to approach the GSE-TW values asymptotically.
Indeed, by plotting the difference from the GSE-TW values against $t$
[\figref{fig:droplet-cumulant}(c)(d) insets],
we find $\braket{H(0,t)}\to\braket{\chi_\mr{GSE}'}$
and $\braket{H(0,t)^2}_\mr{c}\to\braket{\chi_\mr{GSE}^{\prime2}}_\mr{c}$
with finite-time corrections $\sim{}t^{-1/3}$ and $t^{-2/3}$,
respectively.

Moreover, if $q_\mr{L}$ is sufficiently close to $q_\mr{R}$
[e.g., red squares in \figref{fig:droplet-cumulant}(a)(b)],
the data first stay near the curve for $q_{\mathrm{L}}=q_{\mathrm{R}}$,
during which the distribution is
essentially GUE-TW (plus finite-time corrections),
then crossover to the GSE-TW values.
To characterize this crossover,
we tried to collapse the data in \figref{fig:droplet-cumulant}(a)(b)
by rescaling the abscissa in the form $(\Delta{}q)^{\mu}t$,
with $\Delta{}q:=q_{\mathrm{R}}-q_{\mathrm{L}}$ and some exponent $\mu$.
The best collapse was achieved with $\mu=1.4\pm0.1$.
From the theoretical viewpoint, it is reasonable to assume that
this crossover occurs when the height difference
induced by the two different growth speeds, $\Delta{}vt\sim\Delta{}qt$,
becomes comparable to the fluctuation amplitude $(\Gamma{}t)^{1/3}$.
This gives $t\sim(\Delta{}q)^{-3/2}$, hence $\mu=3/2$.
Our data are indeed consistent with this value
 [\figref{fig:droplet-cumulant}(c)(d)].

\begin{figure}[t]
 \centering
 \includegraphics[width=\columnwidth]{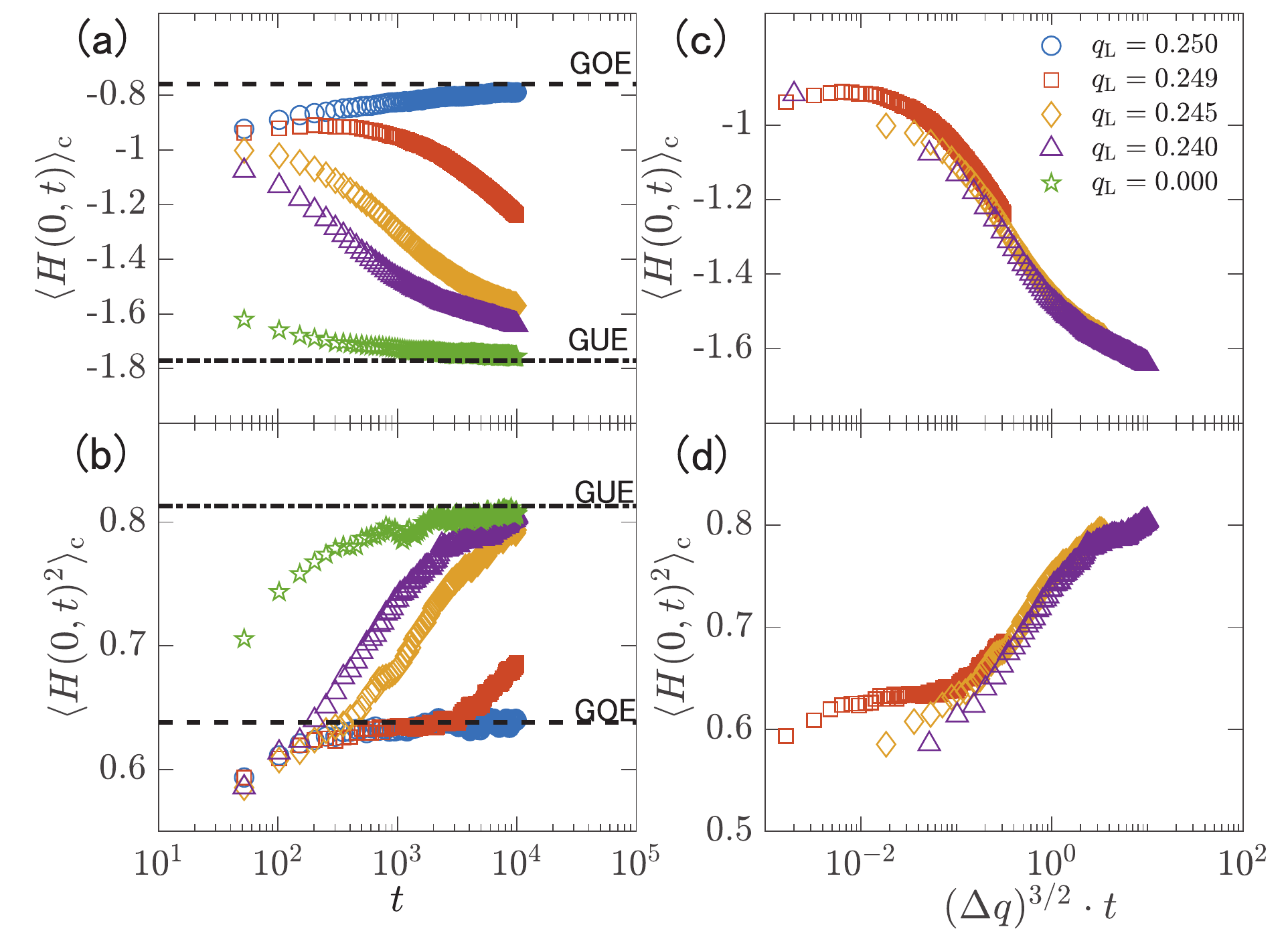}
 \caption{
  The mean and the variance of the rescaled height $H(0,t)$ at the boundary
  for the flat case.
  The horizontal lines indicate the mean and the variance of
  $\chi_\mr{GOE}'$ (dashed) and $\chi_\mr{GUE}$ (dash-dot).
 }
 \label{fig:flat-cumulant}
\end{figure}

\begin{figure}[t]
 \centering
 \includegraphics[width=\columnwidth]{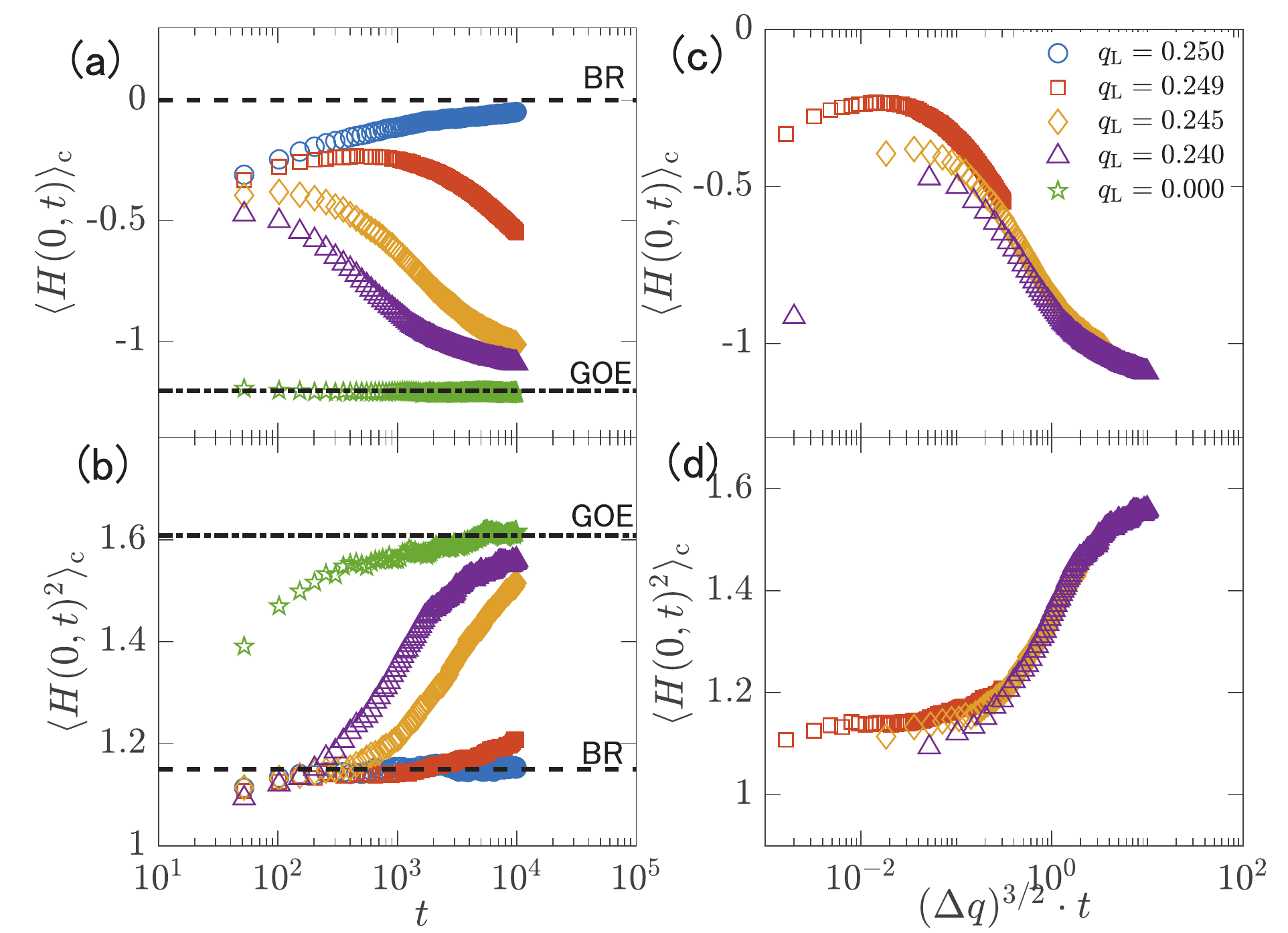}
 \caption{
  The mean and the variance of the rescaled height $H(0,t)$ at the boundary
  for the stationary case.
  The horizontal lines indicate the mean and the variance of
  $\chi_\mr{BR}$ (dashed) and $\chi_\mr{GOE}$ (dash-dot).
 }
 \label{fig:stat-cumulant}
\end{figure}

We also studied the flat and stationary cases
and reached analogous conclusions:
for the flat case (\figref{fig:flat-cumulant})
we find crossover from the GOE-TW to GUE-TW distributions,
and for the stationary case (\figref{fig:stat-cumulant})
from BR to GOE-TW [recall the limiting cases,
\eqsref{eq:limit1} and \pref{eq:limit2}].
The data are found to be consistent
with the same crossover exponent $\mu = 3/2$.

\subsection{Distribution near the Boundary}

\begin{figure}[t]
 \centering
 \includegraphics[width=\columnwidth]{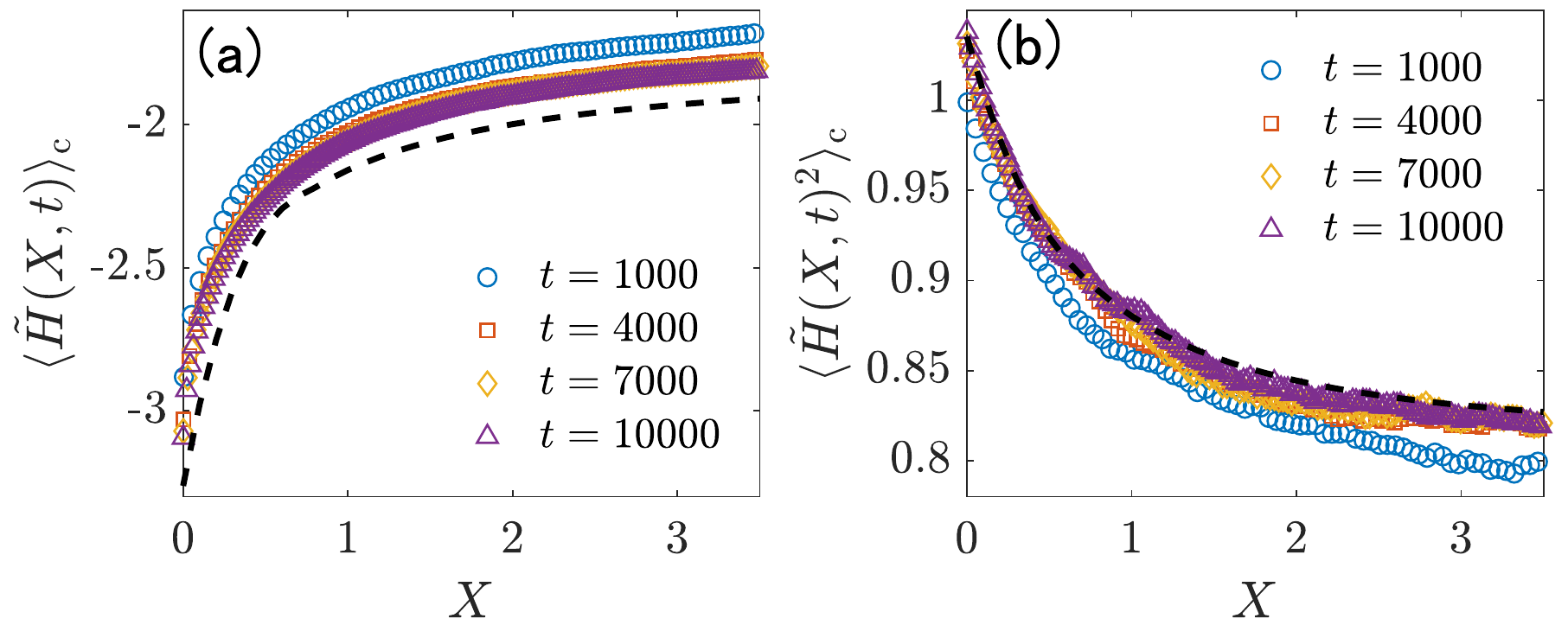}
 \caption{
  The mean and the variance of the compensated rescaled height $\tilde{H}(X,t)$
  near the boundary for the droplet case.
  The dashed lines indicate the theoretical curves
  for the half-space KPZ (Theorem 5.3 of Ref.\cite{Sasamoto2004}),
  numerically evaluated by J. De Nardis and P. Le Doussal.
 }
 \label{fig:onepoint}
\end{figure}

Here we report briefly on the one-point distribution
near the boundary, $X>0$.
In the case of the half-space droplet PNG without boundary nucleation,
Sasamoto and Imamura \cite{Sasamoto2004} derived a formula
(their Theorem 5.3) for the multiple-point joint distribution
of the rescaled height $H(X,\infty)$, or more precisely,
$H(X,\infty)+X^2$ to compensate the parabolic shape
near the top of the droplet [see \eqref{eq:limit1}].
Since this formula also contains information of the one-point distribution
near the boundary, we aim to compare it with our numerical data.
To do so with data obtained at finite times, we need to compensate
not only the parabolic term, but also higher-order nonlinearities
due to the global semicircle shape of the droplet [\eqref{eq:limit-shape}].
This led us to define the compensated rescaled height by
\begin{align}
 \tilde{H}(X,t)
  & :=H(X,t)+\frac{v_{\infty}t}{(1+\sqrt{q})(\Gamma t)^{1/3}}\left[1-\sqrt{1-(x/t)^2}\right] \notag \\
  & \simeq H(X,t) + X^2.                                                                            
\end{align}

Figure~\ref{fig:onepoint} shows the mean and the variance of $\tilde{H}(X,t)$
with $q_\mr{L}=0$ (colored symbols),
compared with the predictions
from Sasamoto and Imamura's formula (dashed lines).
The results for the variance are found to agree with the theoretical curve
 [\figref{fig:onepoint}(b)],
but finite-time corrections are also seen.
Convergence is unfortunately slower for the mean [\figref{fig:onepoint}(a)],
still unreached at the largest time we used.
For the biregional case $q_\mr{L}>0$,
given the result of the crossover time $\sim(\Delta{}q)^{-3/2}$,
convergence to such asymptotic behavior
will be even slower.
However, \figref{fig:onepoint} nicely illustrates
that the one-point distribution in the droplet case crossovers
from the GSE-TW statistics at the origin to the GUE-TW one in the bulk
(far from the boundary).
Analogous crossover, from GUE-TW to GOE-TW for the flat case
and from GOE-TW to BR for the stationary case, is also expected.


\subsection{Spatial Correlation}

\begin{figure*}[t]
 \centering
 \includegraphics[width=0.95\textwidth]{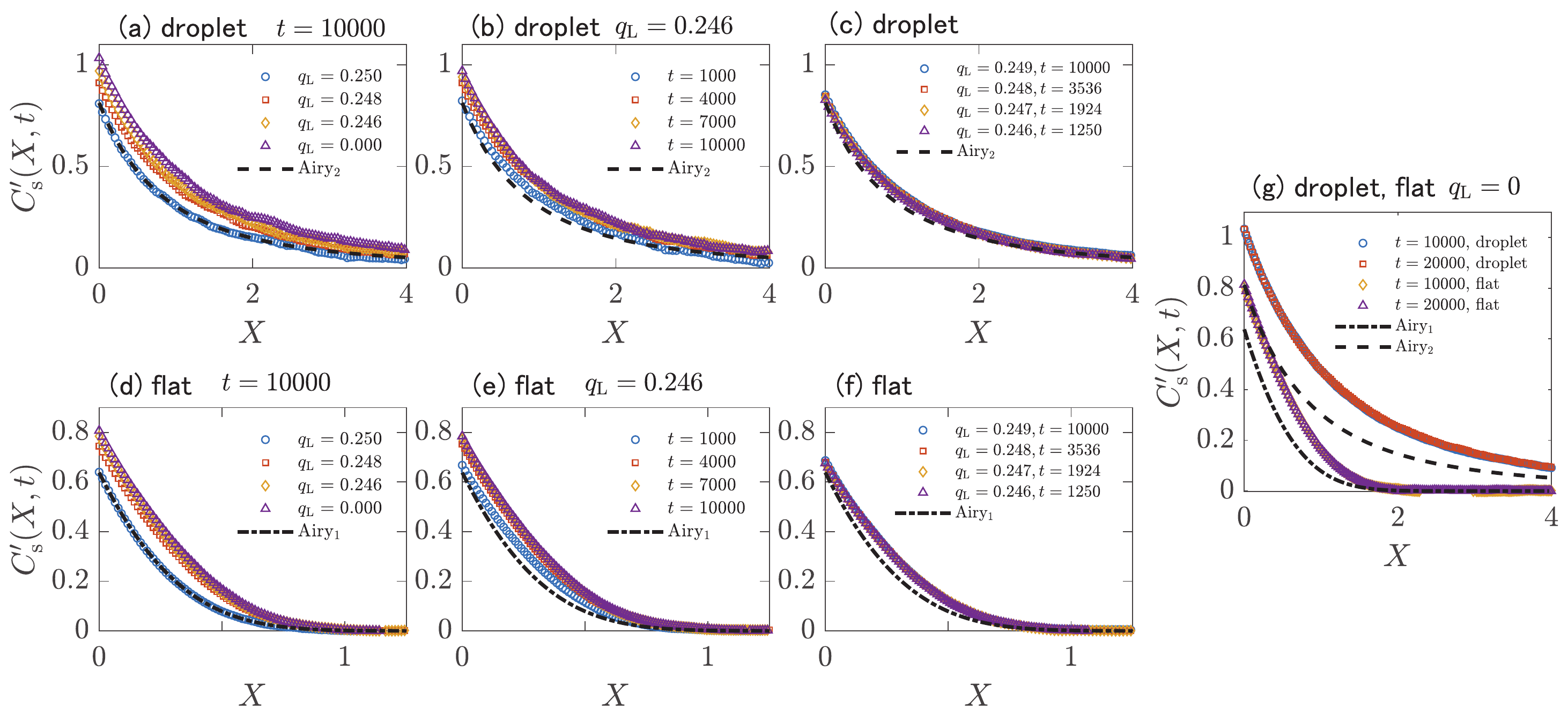}
 \caption{
  The rescaled spatial correlation function $C_{\mr{s}}'(X,t)$
  for the droplet (a-c) and the flat (d-f) cases.
  In (a,d), $t$ is fixed ($t=10000$) and $q_\mr{L}$ is varied.
  In (b,e), $q_{\mr{L}}$ is fixed ($q_{\mr{L}}=0.246$) and $t$ is varied.
  In (c,f), pairs of $q_\mr{L}$ and $t$
  that give the same value of $(\Delta q)^{3/2}t$ are used.
  The panel (g) shows the asymptotic forms of the correlation function
  for the half-space problem, compared with the Airy$_1$ and Airy$_2$
  correlation for the full-space problem.
  The curves for the Airy$_2$ and Airy$_1$ processes were
  numerically evaluated by F. Bornemann \cite{Bornemann-MC2010}.
 }
 \label{fig:correlation}
\end{figure*}

Finally we study the two-point spatial correlation function, defined by
\begin{equation}
 C_{\mathrm{s}}(x,t):=\braket{h(x,t)h(0,t)}-\braket{h(x,t)}\braket{h(0,t)}
\end{equation}
and rescaled as
$C'_{\mr{s}}(X,t):=C_{\mr{s}}(x=Xt^{2/3}/c,t)/(\Gamma t)^{2/3}$.
For the homogeneous growth $q_{\mathrm{L}}=q_{\mathrm{R}}$,
it is known that the asymptotic spatial profile is given by
the stochastic process called the Airy$_{2}$ process for the droplet case
\cite{Praehofer.Spohn-JSP2002}
and the Airy$_{1}$ process for the flat case \cite{sasamoto2005}.
Therefore, the spatial correlation function $C'_{\mr{s}}(X,t)$ is given
directly by their time correlation,
for which analytical formulae are known
\cite{Praehofer.Spohn-JSP2002,sasamoto2005}.
For the half-space droplet case ($q_\mr{L}=0$),
Sasamoto and Imamura's formula \cite{Sasamoto2004} describes this correlation.

Figure~\ref{fig:correlation} shows our numerical results.
For the droplet case, $C'_{\mr{s}}(X,t)$ is plotted
in \figref{fig:correlation}(a) with fixed $t$ and varying $q_\mr{L}$.
We can confirm that the data for $q_\mr{L}=q_\mr{R}=0.250$ are in agreement
with the Airy$_2$ correlation (dashed line).
The corresponding formula by Sasamoto and Imamura is yet to be evaluated,
but since our model with $q_\mr{L}=0$ is equivalent to the half-space PNG
studied by them, we expect that our data
show the functional form of their formula
 [top data set in \figref{fig:correlation}(g)].
For the biregional case $0<q_\mr{L}<q_\mr{R}$, we see the data crossover
from Airy$_2$ to the half-space result, with increasing $\Delta{}q$
(decreasing $q_\mr{L}$) [\figref{fig:correlation}(a)] or increasing $t$
(with fixed $\Delta{}q$) [\figref{fig:correlation}(b)].
This crossover is again controlled by the rescaled time $(\Delta{}q)^{3/2}t$,
which is confirmed in \figref{fig:correlation}(c)
by plotting $C'_{\mr{s}}(X,t)$ for several pairs of $q_\mr{L}$ and $t$
that give the same value of $(\Delta q)^{3/2}t$.
We also tried data collapse of $C'_{\mr{s}}(X,t)$
assuming the combination $(\Delta q)^{\mu}t$ with parameter $\mu$.
It was a difficult task due to unavoidable influence from finite-time effect
and statistical error, but we obtained $\mu=1.3\pm0.2$,
in reasonable agreement with $\mu=3/2$ expected from the theoretical argument
described in \secref{sec:OnePoint}.

The same analysis is carried out in \figref{fig:correlation}(d-f)
for the flat case.
We observe analogous crossover from the Airy$_1$ correlation (dash-dot lines)
to the correlation expected to be that of the half-space flat KPZ problem
[purple triangles in \figref{fig:correlation}(d)
or bottom data set in \figref{fig:correlation}(g)].
To our knowledge, the latter correlation has not been studied theoretically.


\section{concluding remarks}

In this paper, we have proposed a new ``biregional'' situation
for studying the KPZ class, where the interface grows at different speeds
in the left and right halves of space.
We have implemented it using the discrete PNG model
for the three representative geometries,
namely the droplet, flat, and stationary cases,
and numerically studied the fluctuation properties at and near the boundary.
As a result, we have found that they are asymptotically well described
by the half-space problem of the KPZ class, which is characterized
by the sets of the universal statistical properties different from those
for the homogeneous, full-space problem.
In particular, the GSE-TW distribution was found
for the biregional droplet case.
If the growth speed difference is small, we have found crossover
from the full-space statistics to the half-space one,
which is controlled by the rescaled time $(\Delta{}v)^{3/2}t$
with growth-speed difference $\Delta v$.

Our result may also be interpreted in terms of the directed polymer
in random medium,
which provides one of the standard representations of the KPZ class \cite{Kriecherbauer2010,Corwin2012,HalpinHealy.Takeuchi-JSP2015,Takeuchi-a2017}.
In the translation from interface to directed polymer,
growth speed corresponds to the mean depth of the random potential
and the height to the free energy of the polymer,
which tends to find the optimal path under a given random potential.
Now, in our biregional setting, the mean depth of the potential is different
between the two regions.
If this gap is large enough, the optimal path is expected to be found
essentially inside the deeper half space.
The correspondence to the half-space problem is reasonable from this viewpoint.
It is also interesting to recall our finding that the mean interface profile
develops a constant slope near the boundary.
In this sense, a situation similar to imposing the Neumann boundary condition
is spontaneously realized in our setting, providing another explanation
on the correspondence to the half-space problem.
In any case, carrying out direct theoretical analysis
of the biregional KPZ problem is an interesting open problem
left for future studies.

Finally, we believe that our biregional setting
has strong experimental relevance,
compared with the standard half-space problem for which the boundary condition
needs to be controlled.
A study using the liquid-crystal turbulence \cite{PhysRevLett.104.230601}
is ongoing.
We also consider that a similar situation can be realized
in other experimental systems showing KPZ,
such as mutant bacteria colonies \cite{Wakita1997} and
paper combustion \cite{Maunuksela-1997,*Myllys.etal-PRE2001}.
We hope the biregional setting will be a useful platform to investigate
the KPZ half-space problem, both theoretically and experimentally.

\begin{acknowledgments}
 We are indebted to P. Le Doussal, J. De Nardis, and T. Thiery
 for their enlightening suggestions during the whole stage of the work,
 in particular for having encouraged us to study the biregional setting,
 for the arguments based on the noiseless KPZ equation
 \pref{eq:NoiselessKPZ1} and \pref{eq:NoiselessKPZ2},
 and for the numerical evaluation of Sasamoto and Imamura's formula
 presented in \figref{fig:onepoint}.
 We also thank T. Sasamoto and Y. T. Fukai for valuable discussions,
 and F. Bornemann for the theoretical curves of the Airy$_1$ and Airy$_2$
 correlation functions presented in \figref{fig:correlation}.
 This work is supported in part by KAKENHI
 from Japan Society for the Promotion of Science
 (No. JP25103004, JP16H04033, JP16K13846),
 by the grant associated with 2016 Tokyo Tech Challenging Research Award,
 and the National Science Foundation under Grant No. NSF PHY11-25915.
\end{acknowledgments}

\bibliography{reference}

\end{document}